\documentclass[twocolumn,prl]{revtex4}
\usepackage{amssymb}
\usepackage{amsmath}

\newcommand{\half}{\mbox{$\frac{1}{2}$}}
\newcommand{\nuc}[2]{\ensuremath{{}^{#1}{\textrm{#2}}}}
\newcommand{\NOT}{\textsc{not}}

\newcommand{\CNOT}{controlled-\NOT}

\begin{document}
\title{Composite pulses in NMR quantum computation}
\author{Jonathan A. Jones}\email{jonathan.jones@qubit.org}
\affiliation{Centre for Quantum Computation, Clarendon Laboratory,
University of Oxford, Parks Road, OX1~3PU, United Kingdom}
\affiliation{Centre for Advanced ESR,
University of Oxford, South Parks Road, Oxford OX1~3QR, UK}
\date{\today}
\begin{abstract}
I describe the use of techniques based on composite rotations to
combat systematic errors in quantum logic gates. Although developed and
described within the context of Nuclear Magnetic Resonance (NMR)
quantum computing these sequences should be applicable to other
implementations of quantum computation.
\end{abstract}
\maketitle

\section{Introduction}

Quantum computers \cite{Deutsch1985,Bennett2000} are explicitly quantum mechanical systems that use quantum phenomena to perform computational tasks more efficiently than any classical computer.  Unsurprisingly quantum computation has generated enormous interest, reflecting not just its potential technological importance, but also the intellectual importance of the challenge provided to previous formulations of computational complexity theory. This interest is tempered by the apparent difficulty in building large scale devices capable of implementing useful computations, but it has proved fairly simple to build small demonstration devices, and nuclear magnetic resonance (NMR) has played a leading role in this.

The first ideas on how to build quantum computers with NMR \cite{Cory1996,Cory1997,Cory1998a,Gershenfeld1997,Chuang1998b} were swiftly followed by the first implementations of quantum algorithms \cite{Jones1998c,Chuang1998,Chuang1998a,Jones1998d}.  It must be remembered that the great difficulty in preparing NMR systems in pure spin states  has given rise to grave concerns about the direct relevance of NMR techniques to attempts to build large scale devices \cite{Warren1997,Gershenfeld1997a,Jones2000a}, and has even led to questioning of whether many NMR quantum computations can be considered true quantum computations at all \cite{Braunstein1999}.  Despite this, NMR remains an interesting technology for exploring simple quantum phenomena and developing techniques which may find applications in other technologies.  A number of reviews, e.g. \cite{Jones2001a,Vandersypen2004,Heidebrecht2006,Ryan2008,Suter2008}, have described developments in particular areas of NMR quantum computing.  In this paper I will describe the use of composite pulses, a technique developed in conventional NMR \cite{Levitt1986}, to the design of quantum logic gates which are resistant to systematic errors in their implementation \cite{Jones2003a,Cummins2000,Cummins2003,Xiao2006}.


\section{Spin-half nuclei in liquid samples}
There are many possible physical implementations of a qubit, but a particularly natural implementation is provided by a spin-half atomic nucleus, such as \nuc{1}{H}. A two-qubit quantum computer can be built from two atomic nuclei, and so on.  It is necessary that the two nuclei are distinct, so that the the two qubits can be separately addressed, and there must be some sort of spin--spin interaction, so that two-qubit logic gates can be constructed.  This is easily achieved by using two inequivalent nuclei in a molecule.

The spin Hamiltonian \cite{Abragam1961,Slichter1990,EBWbook,Sorensen1983} is in principle quite complicated, but in liquid state (or solution state) samples is greatly simplified by rapid molecular tumbling.  This largely removes intermolecular interactions such as dipolar coupling, and cancels out the anisotropic parts of intramolecular interactions, such as the chemical shift and scalar coupling, reducing them to their isotropic forms.  The cancellation of intermolecular interactions results in an ensemble of identical independent molecules, which can for most practical purposes be treated as a single molecule in a (usually \cite{Anwar2004}) highly mixed spin state.  (This cancellation is not, of course, perfect, and intermolecular and anisotropic interactions remain an important source of spin relaxation.)

An important practical distinction can be made between systems where all the spins are of different nuclear species (a fully heteronuclear spin system) and those with two or more nuclei of the same type (a homonuclear spin system).  As there are a limited number of different spin-half nuclei, among which only six (\nuc{1}{H}, \nuc{13}{C}, \nuc{15}{N}, \nuc{19}{F}, \nuc{29}{Si} and \nuc{31}{P}) have been used in quantum computing experiments so far, it is clear than only very small quantum computers can be fully heteronuclear, but the relative ease of working with such systems makes them popular for implementing simple tasks.

Systems involving quadrupolar nuclei, with spins greater that one half, or nuclei in solid or liquid-crystal samples, have also been studied, but for simplicity in this paper I will only consider systems of spin-half nuclei in the liquid state. The NMR Hamiltonian for such a system is in general given by
\begin{equation}
\mathcal{H/\hbar}=\sum_j \frac{\omega_j}{2}\,\sigma_z^j + \sum_{j<k} \frac{\omega_{jk}}{2}\,\sigma^j\cdot\sigma^k
\end{equation}
although in real systems the coupling strengths $\omega_{jk}$ between distant nuclei can frequently be taken as zero.  Spin--spin couplings are fairly weak, and so in many cases (including all fully heteronuclear systems) it is possible to use the weak coupling approximation
\begin{equation}
|\omega_{jk}|\ll|\omega_j-\omega_k|
\end{equation}
leading to the simplified Hamiltonian
\begin{equation}
\mathcal{H/\hbar}\approx\sum_j \frac{\omega_j}{2}\,\sigma_z^j + \sum_{jk} \frac{\omega_{jk}}{2}\,\sigma^j_z\sigma^k_z
\end{equation}
where the sum is now taken over spin pairs with non-negligible couplings.

NMR texts and papers typically describe spin systems using product operator notation \cite{EBWbook,Sorensen1983}, which is closely related to but not quite identical to conventional physics notation \cite{Jones2001a}.  Within this language a two-spin system would be described by the Hamiltonian
\begin{equation}
\mathcal{H}=2\pi\nu_I\,I_z+2\pi\nu_S\,S_z+\pi{J}\,2I_zS_z
\end{equation}
where
\begin{equation}
I_z=\half\sigma^1_z,\quad S_z=\half\sigma^2_z,
\end{equation}
and the factor of $\hbar$ has simply been dropped by choosing to work in angular frequency units.

\subsection{Quantum logic gates}
It is well known that in order to perform general quantum computations, it is only necessary to implement single-qubit gates, which change the state of a single qubit, and one non-trivial two-qubit gate, for which the final state of at least one of the two qubits involved depends on the initial states of \textit{both} qubits, so that the two-qubit gate encodes some sort of conditional logic \cite{Barenco1995}.

Single-qubit gates correspond to rotating a single spin in its own one-spin Hilbert space. For a one qubit computer, implemented using a single nuclear spin, this can be achieved by applying RF fields.  For simplicity if often best to consider only resonant RF fields, so that the rotation has the form
\begin{equation}\label{eq:pulse}
U(\theta,\phi)=\exp[-\textrm{i}\theta(I_x\cos\phi+I_y\sin\phi)]
\end{equation}
where $\theta$ and $\phi$ are the pulse nutation (flip) and phase angles.  Rotations about axes not in the $xy$ plane can be implemented as sequences of pulses. For example, rotations around the $z$ axis are easily constructed with composite Z-rotations \cite{Freeman1981}, using identities such as
\begin{equation}\label{eq:compZ}
\theta_z=90_y\;\theta_x\;90_{-y}
\end{equation}
where the pulse sequence is written with time running from left to right, so that the leftmost pulse is the first pulse applied.  In larger spin systems it is necessary to do this in a qubit-selective manner; in a fully heteronuclear spin system qubit selection is simple, as every spin will be a long way from resonance with every other spin, and simple pulses applied on resonance can be used, but in homonuclear spin systems more sophisticated approaches are necessary \cite{Jones2001a}.

Next I turn to non-trivial two-qubit gates in two-spin (two-qubit) systems.  In NMR experiments the key two-qubit gate is the controlled-Z gate
\begin{equation}\label{eq:cZ}
\begin{pmatrix}1&0&0&0\\0&1&0&0\\0&0&1&0\\0&0&0&-1\end{pmatrix}
\end{equation}
which is easily converted to a \CNOT\ by a pair of Hadamard gates.  Controlled-Z is symmetric between the two spins and it can be easily decomposed with product operators \cite{Jones1998b} as
\begin{equation}\label{eq:cZpo}
\textrm{controlled-Z}=\exp[-\textrm{i}\,\pi/2\,(\half{E}-I_z-S_z+2I_zS_z)]
\end{equation}
where $E$ indicates the identity matrix.  All four terms commute, and so can be considered individually.  The $\half{E}$ term is just a global phase, and can be ignored as usual.  Terms in $I_z$ and $S_z$ are single qubit rotations, and so can be implemented with single-qubit gates, or simply absorbed into the reference frame \cite{Knill2000,Bowdrey2005}.  This leaves the only important term, which corresponds to evolution under the spin--spin coupling term, $\pi{J}\,2I_zS_z$ for a time $1/2J$.  The spin Hamiltonian will include both Zeeman and coupling terms, but conventional spin-echo sequences \cite{Hahn1950} can be used to remove the undesirable terms \cite{Linden1999a}.

\section{Composite pulses}
Composite pulses \cite{Levitt1986} have found widespread use in conventional NMR experiments to reduce the effects of a wide range of experimental imperfections, most notably off-resonance effects, which arise when the RF field is not quite resonant with the transition so that nutation occurs around a tilted axis, and pulse length errors arising from variations in the strength of the RF field.  (These errors could be better described as \textit{pulse strength errors}, but the unhelpful name is almost universal.)  As similar imperfections are likely to affect most experimental implementations of quantum information processing there has been interest in applying these ideas.

Composite pulses are only one of a whole range of techniques for improving the quality of RF excitation in NMR experiments.  Traditionally these can be divided into composite pulses, made up of a small number of pulses, each with the same frequency and strength but differing in length and phase, and shaped pulses \cite{Freeman1998}, which contain a very large number of elements, each with the same frequency and length but differing in strength and phase.  A further pragmatic distinction is that composite pulses can often be derived and explained using analytical approaches, while shaped pulses are frequently derived numerically using analytic ideas only as an outline guide.  These distinctions have increasingly broken down with the advent of strongly modulated composite pulses \cite{Fortunato2002}: these contain a small number of pulse elements, but these pulses are permitted to vary in frequency and amplitude as well as phase, and the pulse sequence is obtained by numerical optimisation.  More recently still these have largely been superseded by arbitrary shaped pulses developed using optimal control theory, usually based on the GRAPE algorithm \cite{Schulte-Herbruggen2005,Khaneja2005}.  In this review I will confine myself to ``conventional'' composite pulses, with analytical derivations.

Most conventional composite pulses are not suitable for use in quantum computers, as they are optimised for particular classes of initial state: for example, most composite $180^\circ$ pulses are optimised either for inverting the population of a spin or for producing a spin echo.  By contrast, pulses used on quantum computers must be general rotors, which perform well for any initial state. Composite pulses of this kind are rarely used in conventional NMR, but a small number of so-called Class~A composite pulses \cite{Levitt1986} are known, and these have been developed for use in quantum computation.  A method for constructing general rotors from conventional point-to-point pulses has also been described \cite{Luy2005}.

\subsection{Fidelity Measures}
The quality of a composite pulse for quantum computing can be assessed in various ways, but in practice there are two important families of approaches.  The most direct approach is to expand the propagator for the composite pulse as a power series in the size of the error, and determine the size and order of the lowest order error term.  As an example consider implementing a $180^\circ_x$ rotation using a naive pulse with a fractional pulse length error of $\epsilon$, so that the flip angle of the pulse is in fact $180\times(1+\epsilon)$.  The ideal propagator is then
\begin{equation}
U=\exp[-\textrm{i}\,\pi\,\sigma_x/2]=
\begin{pmatrix}0&-\textrm{i}\\-\textrm{i}&0\end{pmatrix}
\end{equation}
while the actual propagator is
\begin{align}
V&=\exp[-\textrm{i}\,\pi(1+\epsilon)\,\sigma_x/2]\\
&=
\begin{pmatrix}0&-\textrm{i}\\-\textrm{i}&0\end{pmatrix}
-\epsilon\begin{pmatrix}\pi/2&0\\0&\pi/2\end{pmatrix}+\textrm{O}(\epsilon^2)
\end{align}
and so the naive pulse has an error of order $\epsilon$.  Alternatively the quality can be assessed by calculating the propagator fidelity between $U$ and $V$, given by
\begin{equation}
F=|\textrm{Tr}(VU^{-1})|/\textrm{Tr}(UU^{-1}),
\end{equation}
and then expanding the fidelity as a power series in the error.  For the naive pulse considered above the fidelity is
\begin{equation}
F=1-\epsilon^2\pi^2/8+\textrm{O}(\epsilon^4)
\end{equation}
and the naive pulse has \textit{infidelity} of order $\epsilon^2$.  The difference between these two methods of assessing a pulse must be borne in mind when comparing pulses in different papers; in general an error of order $n$ will correspond to an infidelity of order $2n$ \cite{Brown2004,Brown2005}.

\subsection{Off-resonance errors}
An early composite $90^\circ$ pulse tackling off-resonance errors was described by Tycko \cite{Tycko1985}, replacing a $90^\circ_x$ pulse with the three pulse sequence $385^\circ_x320^\circ_{-x}25^\circ_x$.  This has subsequently been generalised to give the \textsc{corpse} family of composite pulses \cite{Cummins2000,Cummins2003}, in which a $\theta_x$ pulse is replaced by three pulses, applied along the $+x$, $-x$ and $+x$ axes as before, with flip angles given by
\begin{align}
\theta_1&=2n_1\pi+\frac{\theta}{2}-\arcsin\left(\frac{\sin(\theta/2)}{2}\right)\\
\theta_2&=2n_2\pi-2\arcsin\left(\frac{\sin(\theta/2)}{2}\right)\\
\theta_3&=2n_3\pi+\frac{\theta}{2}-\arcsin\left(\frac{\sin(\theta/2)}{2}\right)
\end{align}
where $n_1$, $n_2$ and $n_3$ are integers, with the best results  \cite{Cummins2003} occurring for $n_1=n_2=1$ and $n_3=0$.  These sequences have been demonstrated by NMR \cite{Cummins2000}, \textsc{squid} \cite{Collin2004} and neutral atom \cite{Rakreungdet2009} experiments.  Pulse sequences have also been designed which are tailored for particular off-resonance effects \cite{Cummins2001}.

\subsection{Pulse length errors}
While off-resonance errors are important in conventional NMR, they can largely be avoided in quantum information processing experiments.  Pulse length errors, however, remain a universal problem, arising from inhomogeneities in the RF field, either in space (over a macroscopic sample) or in time (due to slow fluctuations in amplifier power).  There has, therefore, been considerable interest in composite pulses to tackle pulse length errors, which can largely be traced back to a three pulse composite $180^\circ$ pulse due to Tycko \cite{Tycko1985} or to the BB1 family of sequences discovered by Wimperis \cite{Wimperis1994}.  Tycko's pulse sequence has been generalised to give the \textsc{scrofulous} family of composite pulses \cite{Cummins2003}, but using the BB1 family is preferable in most cases.

BB1 differs from many other composite pulses in that it seeks to design an error-correcting pulse, which can be combined with the naive error-prone pulse to give a more accurate compound pulse, much as a contact lens can be used to correct eyesight.  Originally \cite{Wimperis1994} this error correcting sequence (sometimes called a W1 sequence) was placed before the naive pulse, but it can instead be placed after the naive pulse, or indeed in the middle of it \cite{Cummins2003,Xiao2006}.  It comprises three pulses, in the form $180^\circ_{\phi_1}\,360^\circ_{\phi_2}\,180^\circ_{\phi_1}$, with $\phi_2=3\,\phi_1$ and
\begin{equation}
\phi_1=\pm\arccos\left(-\frac{\theta}{4\pi}\right)
\end{equation}
where the choice of sign is unimportant as long as it is made consistently.

In addition to NMR experiments \cite{Xiao2005,Xiao2006} BB1 pulses have been demonstrated in electron spin resonance \cite{Morton2005a,Ardavan2007a} and neutral atoms \cite{Rakreungdet2009}, and have inspired applications in other fields \cite{Wesenberg2003,Ardavan2007b,Testolin2007}.

\subsection{Higher precision sequences}
BB1 has proved a remarkably successful composite pulse, and is surprisingly difficult to improve upon.
BB1 pulses can be derived by designing composite pulses which suppress first order pulse length errors, but it turns out that BB1 also suppresses second order errors automatically, leaving only third order errors (sixth order infidelity).  It is not clear why this fortuitous double cancellation occurs, and it is not a general feature of composite pulses.  Other pulses with similar properties are known \cite{McHugh2005}, but these have no advantages over BB1.  Beyond this, BB1 pulses are also relatively robust to off resonance-errors \cite{Cummins2003}, and generally insensitive to small errors in their implementation, so that BB1 pulses work in practice very much as expected from theory \cite{Xiao2006}.

Although BB1 has proved highly successful, it is obviously interesting to seek still better pulse sequences, and Brown \textit{et al.} have tackled this in two ways \cite{Brown2004,Brown2005}.  Firstly they have shown how the BB1 approach can, in effect, be nested, creating ever higher orders of simultaneous correction.  A robust $90^\circ$ pulse from the B4 family of pulses (which remove the third order error term) has been implemented in NMR experiments \cite{Xiao2006}, but this composite pulse is very long (the correction sequence contains 27 pulses with a total length equivalent to a $7200^\circ$ rotation) and does not perform much better than BB1.  Secondly they have described a general method, using insights from the Solovay--Kitaev theorem \cite{Dawson2006}, to show how arbitrarily accurate composite pulses can be constructed in general, by building a series of correction sequences which correct errors one order at a time.  An expanded version of part of their method written in more conventional NMR notation is also available \cite{Alway2007}.  Once again high order corrections sequences developed using these ideas can become extremely long, and it is not clear how well such complex pulses will work in practice.  In comparison BB1 pulses are extremely robust \cite{Xiao2006}, and so may prove the ideal compromise between theoretical precision and practical implementation.

\subsection{Two qubit gates}
The method can be extended to build two-qubit gates which are robust to variations in the size of the underlying scalar coupling \cite{Jones2003b,Jones2003} using an analogy between rotations on the Bloch sphere and the rotations in a multi-qubit Hilbert space which correspond to evolution under the spin--spin coupling.  In combination with the single qubit gates described previously these provide a universal set of robust quantum logic gates.  They have been demonstrated using NMR techniques \cite{Xiao2006}, but it is not yet clear how important they will prove in practice.

\subsection{Conclusions}
Composite pulse techniques adapted from conventional NMR experiments have already proved to be extremely useful in NMR quantum computation, and have begun to find wider applications in related fields.  Pulse length errors, which ultimately relate to uncertainties in the strength of the external control fields, are likely to have analogies in many implementations of quantum information processing, and the BB1 pulse sequence provides an apparently ideal method for tackling these, combining good suppression of errors, relative simplicity, and an apparent robustness to imperfections in its implementation.

\begin{acknowledgments}
I thank the UK EPSRC for financial support.
\end{acknowledgments}

\end{document}